# Study of gauge (in)dependence of monopole dynamics

Shinji Ejiri [a], Shun-ichi Kitahara [a], Yoshimi Matsubara [b], Tsuyoshi Okude [a], Tsuneo Suzuki [a] and Koji Yasuta [a]

[a]Department of Physics, Kanazawa University, Kanazawa 920-11, Japan

[b]Nanao Junior College, Nanao, Ishikawa 926, Japan

We investigated the gauge (in)dependence of the confinement mechanism due to monopole condensation in SU(2) lattice QCD by various abelian projections. We found (1) the string tension can be reproduced by monopoles alone also in Polyakov gauge and (2) the behaviors of the Polyakov loop at the critical temperature seem to be explained by the uniformity breaking of the monopole currents in every gauge.

## 1. Introduction

Many people believe that the quark confinement mechanism in QCD can be understood as dual Meissner effect due to abelian monopole condensation. And this picture is supported by recent Monte-Carlo simulations of abelian projected QCD in maximally abelian gauge (MA gauge). There are infinite ways of abelian projection extracting an abelian component. If abelian monopole condensation is really the confinement mechanism in QCD, it must be independent of the way of abelian projection, i.e., gauge independent. The purpose of this report is to study monopole physics in gauges other than the MA gauge such as Polyakov gauge.

The method of our studies are the followings. (1) We perform usual Monte-Carlo simulations of QCD. (2) Choosing a gauge, we extract an abelian theory by abelian projection [1, 2]. (3) We measure observables written in terms of abelian link variable. When the features of QCD can be reproduced by such an abelian variable, we call this property abelian dominance. (4) The operators of the abelian observables ( Wilson loop, Polyakov loop, $\langle\bar{\psi}\psi\rangle, \ldots$ ) can be decomposed into contributions from monopole and from photon. We measure these two parts separately. When the features of QCD can be explained by the contribution from monopole alone, it is called monopole dominance.

Many results in the MA gauge and a few results in the other gauges about abelian dominance and monopole dominance have been reported.

| Results in MA gauge | |
|---|---|
| string tension $\sigma$ | '90 Suzuki-Yotsuyanagi[3], '91 Hioki et al.[4], '92 Suzuki[5], '94 Shiba-Suzuki[6], '94 Ejiri et al.[7], |
| Polyakov loop $\langle P\rangle$ and $T_c$ | '91 Hioki et al.[4], '92 Suzuki[5], '94 Suzuki et al.[8], |
| quark condensate $\langle\bar{\psi}\psi\rangle$ | '94 Miyamura[9], |
| pion mass $m_\pi$, $\rho$ mass $m_\rho$ | '95 Miyamura-Origuchi[10], '95 Kitahara et al., |
| instanton charge | '95 Miyamura-Origuchi[10], |

| Results in the other gauges | |
|---|---|
| string tension (Polyakov gauge) | '95 Ejiri et al., (This report) |
| Polyakov loop | '94 Suzuki et al.[8], |

## 2. Study in Polyakov gauge

We consider finite temperature QCD. The abelian dominance of the quantities derived from the Polyakov loops is trivial in the case of the Polyakov gauge.

We define a matrix $P(s)$ as follows

$$P(s) = \prod_{i=1}^{N_4} U(s + (i-1)\hat{4}, 4). \qquad (1)$$

The Polyakov gauge is defined by diagonalizing $P(s)$. And an abelian link variable $u(s,\mu)$ is extracted as $U(s,\mu) = c(s,\mu)u(s,\mu)$, where $u(s,\mu)$ is diagonal. Then all $U(s,4)$ are diagonal and abelian. $U(s,4) = u(s,4)$. It means that the full Polyakov loop is equal to the abelian Polyakov loop. The Fadeev-Popov determinant of the Polyakov gauge is given in a simple form [11].

The abelian Polyakov loop operator:

$$P = \exp(i\sum_{i=1}^{N_4} J_4(s+(i-1)\hat{4})\theta_4(s+(i-1)\hat{4}) \quad (2)$$

$$U(s,4) = u(s,4) = \mathrm{diag}\{e^{i\theta_4(s)}, e^{-i\theta_4(s)}\}$$

can be decomposed into a photon part and a Dirac string part using following identities:

$$\theta_4(s) = -\sum_{s'} D(s-s')[\partial'_\nu \theta_{\nu 4}(s') + \partial_4(\partial'_\nu \theta_\nu(s'))], \quad (3)$$

where $\theta_{\mu\nu} = \partial_\mu \theta_\nu - \partial_\nu \theta_\mu$. $D(s-s')$ is the lattice Coulomb propagator. The current conservation $\partial'_4 J_4(s) = 0$ leads us to [8]

$$P = P_1(\text{photon part}) \cdot P_2(\text{Dirac string part}), \quad (4)$$

$$P_1 = \exp\{-i\sum_{i=1}^{N_4} J_4(s+(i-1)\hat{4}) \sum_{s'} D(s+(i-1)\hat{4}-s')\partial'_\nu \bar{\theta}_{\nu 4}(s')\}, \quad (5)$$

$$P_2 = \exp\{-2\pi i\sum_{i=1}^{N_4} J_4(s+(i-1)\hat{4}) \sum_{s'} D(s+(i-1)\hat{4}-s')\partial'_\nu n_{\nu 4}(s')\}, \quad (6)$$

where

$$\theta_{\mu\nu}(s) = \bar{\theta}_{\mu\nu}(s) + 2\pi n_{\mu\nu}(s), \quad (-\pi < \bar{\theta}_{\mu\nu}(s) \leq \pi).$$

Here ${}^*n_{\mu\nu}(s) = \frac{1}{2}\epsilon_{\mu\nu\rho\sigma} n_{\rho\sigma}(s)$ is the Dirac string. The monopole currents $k_\mu(s)$ are the boundary of the Dirac string: $k_\mu = \partial_\nu\, {}^*n_{\nu\mu}$ [12].

The string tension is derived from the correlation of Polyakov loops. The simulations are performed on a $24^3 \times 4$ lattice. The measurements are done every 50 sweeps after 2000 sweeps for thermalization. We used 100 configurations for the Polyakov loop and 500 or 1000 configurations for Polyakov loop correlation. The Polyakov loop is plotted in Fig. 1, and a typical Polyakov loop correlation ($\beta = 2.27$) is shown in Fig. 2. The string tension was obtained from Polyakov loop correlation by least square fit assuming that the static quark anti-quark potential is given by linear + Coulomb + constant terms. The $\beta$ dependence of the string tension is shown in Fig. 3.

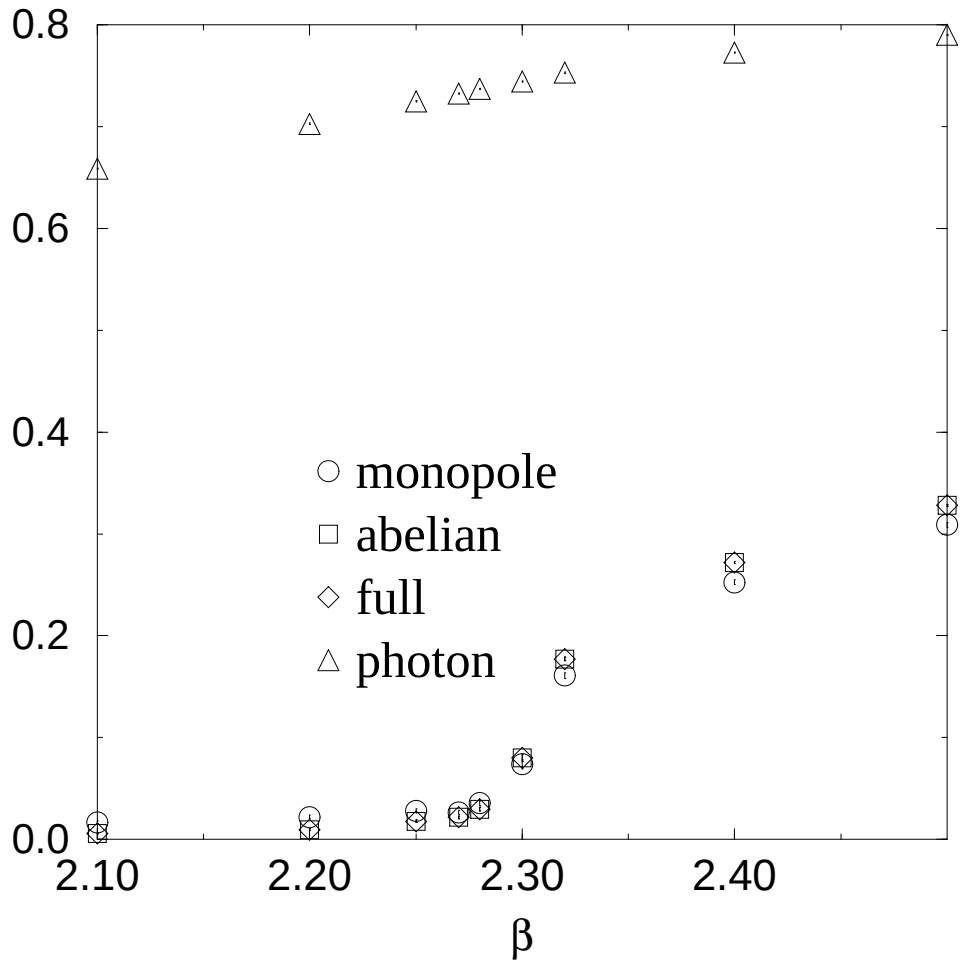

Figure 1. Monopole (Dirac string) and photon contributions to Polyakov loops in the Polyakov gauge

We can see in Fig. 1 that the abelian flux is squeezed by the Dirac string also in the Polyakov gauge.

In the Polyakov gauge, we found that the abelian Polyakov loop correlations give the same string tension as the full one, (abelian dominance) and that the Polyakov loop correlations from monopoles reproduce much the same string tension near the scaling region. (monopole dominance)

### 3. Monopole dynamics at $\beta_c$

Monopole dominance of Polyakov loops was shown in various gauges [8]. It suggests that the monopole dynamics which makes these behaviors of Polyakov loop at $\beta_c$ is gauge independent.

Considering $J_4(s) = 1$ along the Polyakov line, the monopole contribution to Polyakov loop can

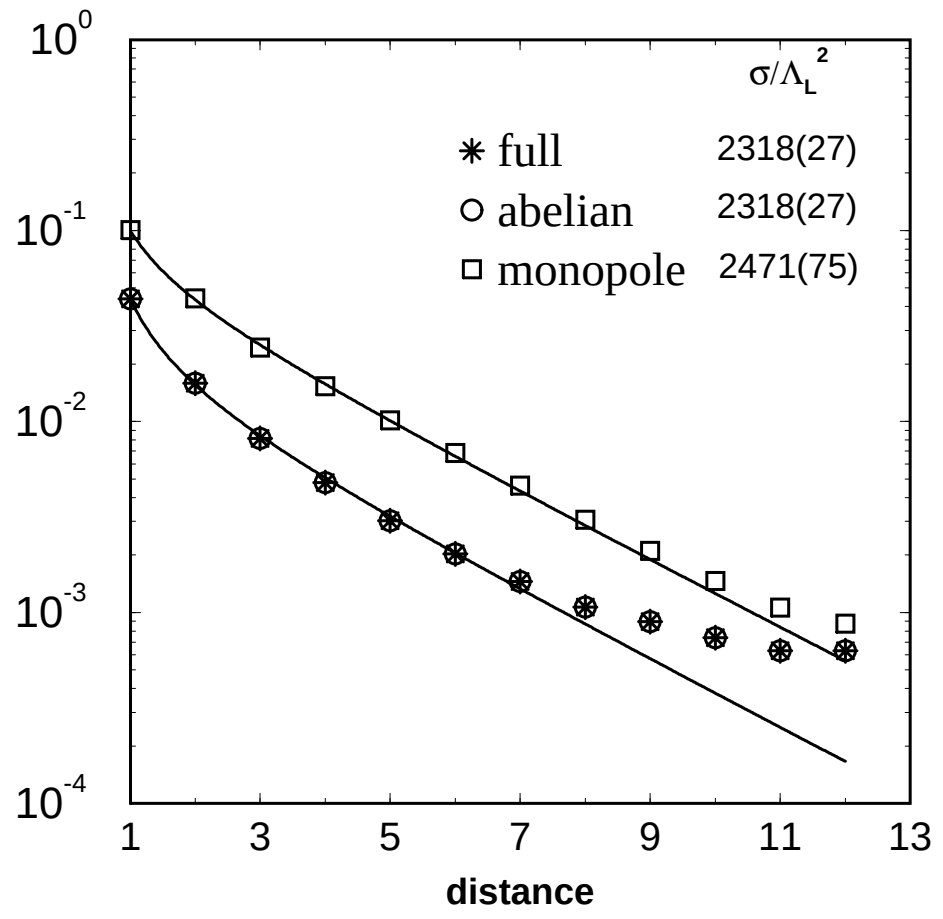

Figure 2. Polyakov loop correlations in the Polyakov gauge

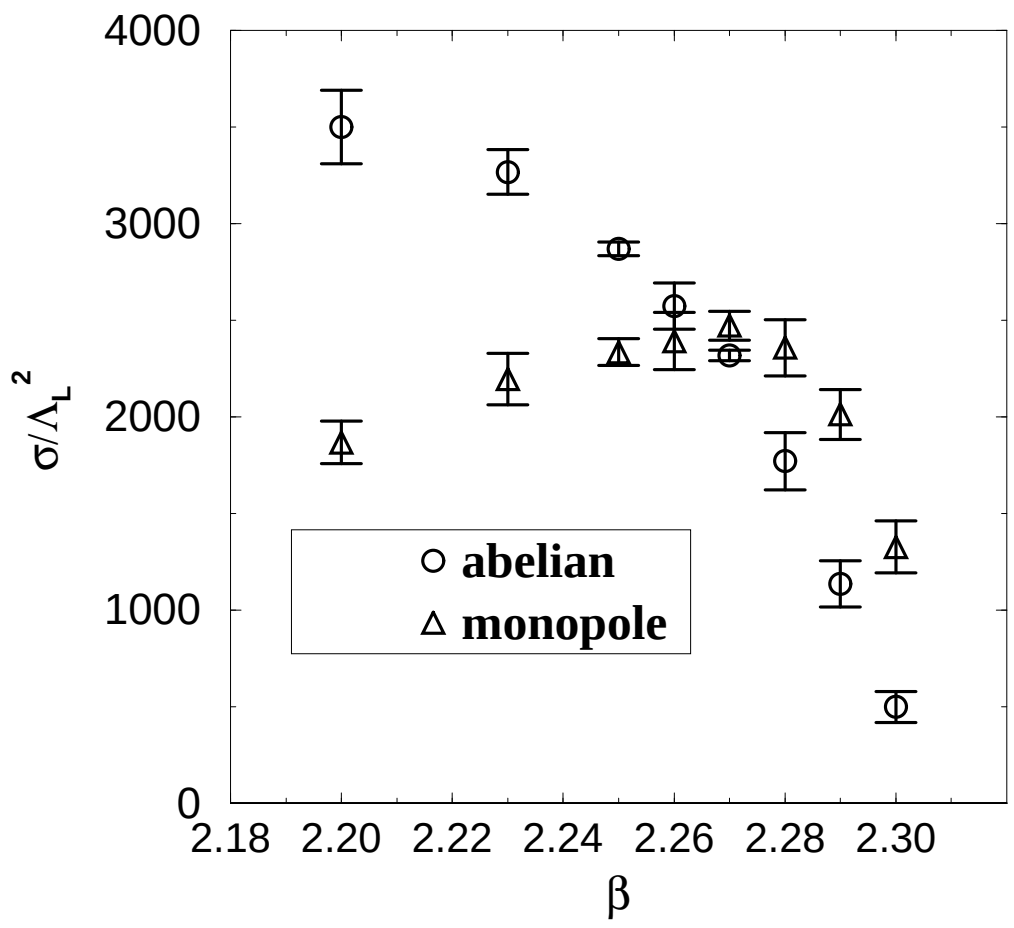

Figure 3. $\beta$ dependence of the string tension in the Polyakov gauge

be rewritten by

$$\langle P\rangle = \langle \exp(-2\pi i \sum_{\vec{s}'} D_3(\vec{s}-\vec{s}')\frac{1}{2}\epsilon_{\nu 4\rho\sigma}\partial'_\nu {}^*\!n_{\rho\sigma}(\vec{s}'))\rangle. \quad (7)$$

Here $D_3(\vec{s})$ is the 3-dimensional Coulomb propagator and ${}^*\!n_{\mu\nu}$ is the Dirac string projected in the 3-dimensional space,

$$ {}^*\!n_{\mu\nu}(\vec{s}) = \frac{1}{2}\epsilon_{\mu\nu\rho\sigma}\sum_{s_4} n_{\rho\sigma}(s). \quad (8)$$

This reduces to

$$\langle P\rangle = \langle \exp(2\pi i \frac{\Omega(\vec{s})}{4\pi})\rangle, \quad (9)$$

where $\Omega(\vec{s})$ is the solid angle made by monopole loops projected in the 3-dimensional space from the Polyakov loop $(\vec{s})$.

Notice that this equation shows that the value of Polyakov loop depends only on monopole currents and does not depend on a form of Dirac sheet.

The behaviors of the monopole loops were studied in SU(2) QCD in the MA gauge [13]. In the confinement phase, there are one long connected monopole loop which distributes uniformly and some short loops. The length of the long loop becomes shorter as $\beta$ becomes larger. In deconfinement phase, a long loop disappears.

The properties of Polyakov loop can be understood by these monopole dynamics qualitatively. In the confinement phase ($T < T_c$), long monopole loops distribute uniformly. The $\Omega$ can be random from 0 to $4\pi$. The value of the Polyakov loop at each point is random. The average of the Polyakov loop is zero. On the other hand, in the deconfinement phase ($T > T_c$), there is a space where no monopole exists. In such a space, $\Omega$ takes small value and the local Polyakov loops are nearly one. The average of the Polyakov loop becomes non-zero.

This picture seems to be correct in the MA gauge. In other gauges, the monopoles are so dense that we can not see the breaking of uniformity clearly.

Hence to check if this picture is correct in other gauges, we investigated the histogram of the solid angle. Monte-Carlo simulations were performed on $24^3 \times 4$ lattices at $\beta = 2.2, 2.3$ and $2.4$, $\beta = 2.3$ is critical $\beta$. We took 50 configurations

In Fig. 4, the shape of the histogram changes at $\beta_c$ in every gauge. It is almost flat in the confinement phase, whereas it has a peek near zero in the deconfinement phase.

Notice that, in the Polyakov gauge, there are several methods of diagonalizing $\prod_{s_4} U_4$. If we diagonalize it with a condition $[\prod U_4]_3 > 0$, the histogram becomes antisymmetric. So we used the method of diagonalizing it without such a condition.

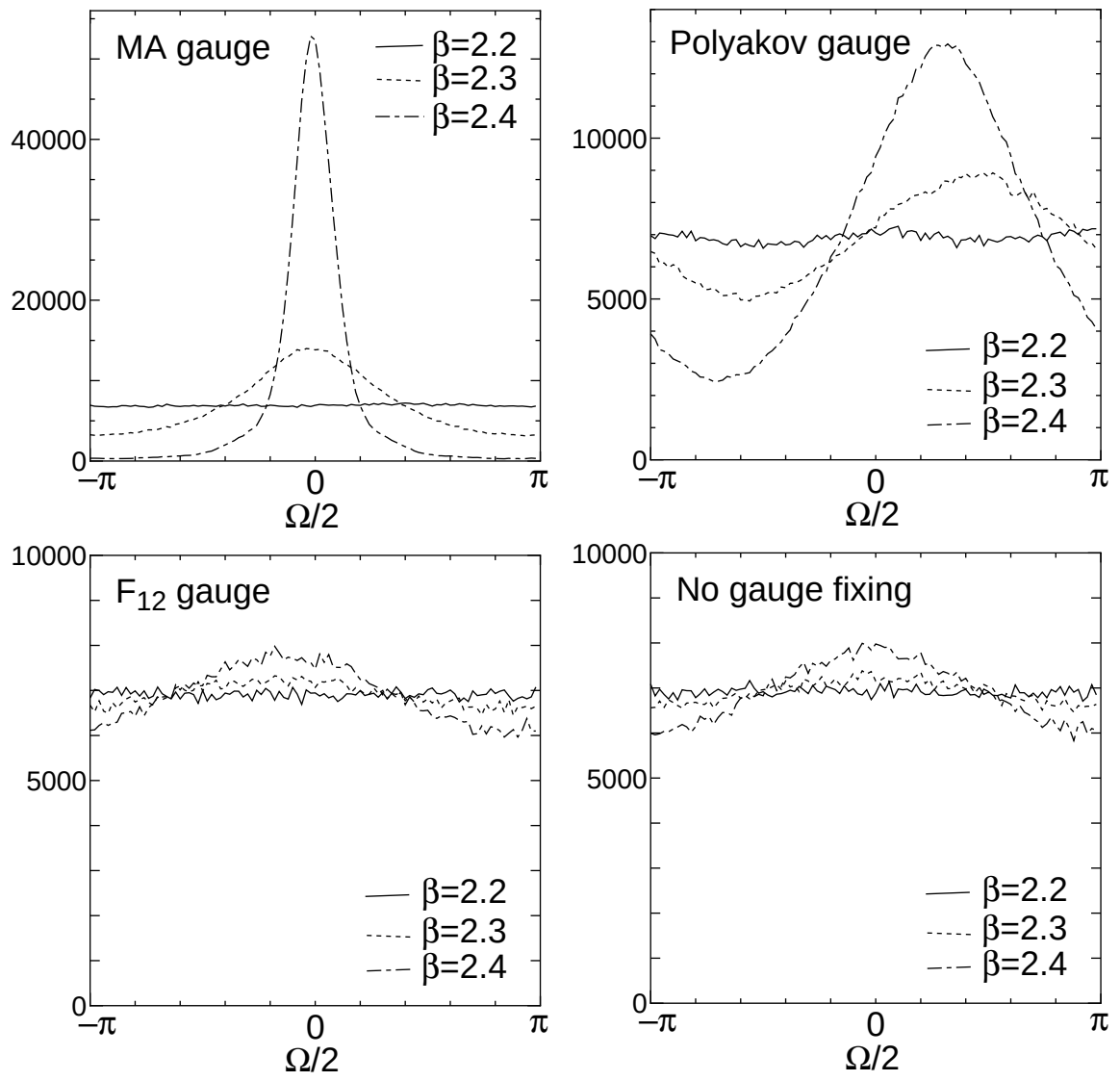

Figure 4. Histograms of $\Omega/2$ on $24^3 \times 4$ lattices with 50 configurations.

## 4. Summary and Discussion

We found the following results. (1) The abelian dominance of the quantities derived from the Polyakov loops is exact in the Polyakov gauge. (2) The string tension and the characteristic behaviors of Polyakov loops are reproduced by monopoles also in the Polyakov gauge. (3) The properties of Polyakov loops have a close relation to the uniformity breaking of the monopole loops at $T_c$, and this picture seems to be gauge invariant.

In Fig. 3 some difference between the values of the sting tension from monopoles and full one are seen at the strong coupling region in the Polyakov gauge. The study of finite size scaling in the $T = 0$ system is important and in progress. We must measure the string tension using Wilson loops on larger lattice.

In the Polyakov gauge, there are several ways of diagonalizing $\prod_{s_4} U_4$, and the monopole currents seem to depend on this procedure. We must study if these dependence affect physical quantities.

In the study of the relation of the monopole and the string tension in the MA gauge [7], it was found that not all monopole currents contribute to the string tension. Only a long monopole loop contributes to the string tension, and the other short monopole loops do not contribute. In the other gauges, the monopole density is so dense that we can not extract the monopole currents which contribute to the confinement. It seems that there are many monopole currents which do not contribute to the confinement in these gauges. We think that these are the reasons why the histograms in the gauges other than the MA gauge do not show so clear change, and that these results show that there exist the monopole currents which are responsible for the confinement. The distributions of these monopole currents change at critical temperature in every gauge.